\documentclass{article}
\usepackage{amsmath}

\usepackage[latin1]{inputenc}
\usepackage[T1]{fontenc}
\usepackage[document]{ragged2e}
\usepackage{ulem}
\usepackage{color}
\usepackage{graphicx}
\newcommand{\add}[1]{#1}
\newcommand{\del}[1]{}

\begin{document}

\noindent {\large \bf Analysis of point defects in graphene using low dose scanning transmission electron microscopy imaging and maximum likelihood reconstruction}

\bigskip

\centering
Christian~Kramberger\textsuperscript{\textsf{1}},
Andreas~Mittelberger \textsuperscript{\textsf{1}},\\
Christoph~Hofer \textsuperscript{\textsf{1}},
Jannik C.~Meyer\textsuperscript{\textsf{1}}

{\it%
\textsuperscript{1}\,Faculty of Physics, University of Vienna, Boltzmanngasse 5, 1090 Vienna, Austria
}

\justify
\bigskip
{\bf abstract:} Freestanding graphene displays an outstanding resilience to electron irradiation at low electron energies. Point defects in graphene are, however, subject to beam driven dynamics. This means that high resolution micrographs of point defects, which usually require a high electron irradiation dose might not represent the intrinsic defect population. Here, we capture the inital defects formed by ejecting carbon atoms under electron irradiation, by imaging with very low doses and subsequent reconstruction of the frequently occuring defects via a maximum likelihood algorithm. 

\bigskip
\noindent{\bf date:} \today

\section{Introduction}
With the advent of aberration correction \cite{Haider1998,Krivanek1999}, electron microscopy has entered the realm of atomic resolution for light elements \cite{Suenaga2007,Meyer2008c,Krivanek2010d}. While this in principle would enable studies on individual organic molecules, beam damage is posing a serious challenge in this field \cite{Egerton2012,Hovden2012}. The structures of interest are destroyed or altered by the probing electron beam, before they could possibly be imaged with satisfactory statistics. A prime example of dynamic entities under electron irradiation are vacancy defects in graphene. Defects in graphene are typically more beam-stable than organic molecules, nevertheless they change configuration under the typical doses that are needed for a high resolution image \cite{Kotakoski2011,Warner2012b}. To date, the statistics on divacancy reconstructions have shown a non-thermal inverted population for the three basic divacancies with 0, 3 and 4 pairs of a 5 and a 7 membered ring \cite{Boerner2016,kotakoski2014imaging}. Here we demonstrate the application of the maximum likelihood reconstruction \cite{meyer2014atomic,Kramberger2016} to very low-dose micrographs containing sparse and not directly visible dynamically created point defects. The population of point defects is consistent with single and double atom ejection events dominating over sequential Stone-Wales transformations. Only one type of divacancy is observed while the other divacancies appear to be absent. 
 
\section{Overview}

This paper demonstrates the recovery of point-defect images in graphene based on scanning transmission electron microscopy (STEM) data recorded at 1-2 orders of magnitude lower doses than typically used for direct imaging of the atomic structure. \add{In similarity to cryo-electron microscopic studies of organic biological molecules \cite{Frank2006,Zhou2008}, we aim to distribute the dose over many identical copies of an object, and to retrieve the equivalent of a high-dose image via a suited reconstruction algorithm.  However, in contrast to large organic molecules, the location of individual point defects (a few missing atoms) can not be discerned in the individual low-dose exposures (e.g. Figs\@.~\ref{raw_map}\&\ref{raw_fft}). Hence, any algorithm that requires to locate and classify individual objects prior to averaging (as is common in the single-particle analysis of biological molecules) will not work for this purpose.
  
  To solve this problem, we have developed a new algorithm that can recover repeatedly occuring deviations from the periodic lattice even if they can not be located in the noisy images \cite{meyer2014atomic,Kramberger2016}. The algorithm employs a maximum-likelihood approach where a set of model images is iteratively optimized so that the models predict the experimental data as well as possible. Apart from the underlying periodic lattice and an approximately expected lateral extension of the defect structures, no a-priori assumptions on the defect structures are required.  Ref. \cite{meyer2014atomic} contained a general description of the ideas in context with other literature, and a first proof of concept based on simulations. In Ref. \cite{Kramberger2016}, we explored the limitations of the new approach, and optimized the implementation to be applicable with real experimental data (shown with high-dose images that were artificially resampled to mimic low-dose conditions). In order to obtain true low-dose images, pristine areas of the sample have to imaged, without pre-exposing the respective sample area for focusing: This is achieved by an automated low-dose acquisition scheme, where focus references are taken at the corners of the region of interest and the focus is then interpolated (this is reported elsewhere \cite{Mittelberger2017}).

  The present paper shows a reconstruction from true low-dose data, expanding on those aspects that are important with real experimental data (and have not been covered in previous simulation-based works) along with a discussion of the findings. In short, these aspects are as follows: After data acquisition, the underlying periodic lattice of all micrographs has to be un-distorted and brought into registry. Then the probability distribution function (PDF) can be collected for different expectation values. With this information the likelihood that the entire experimental dataset is an actual observation of any composition of a finite set of small repeating patches can be calculated. The expression of the likelihood is given in Eqs\@.~1\&2 and the maximum likelihood algorithm optimizes the likelihood value by iteratively adjusting a set of model images. In the end, effective high-dose images, as shown in Figure~\ref{reconstruction}, of the most frequently occuring defects are obtained, along with relative weights representing their density.}

\section{Experimental}

\begin{figure}[htb]%
\includegraphics*[width=0.5\linewidth]{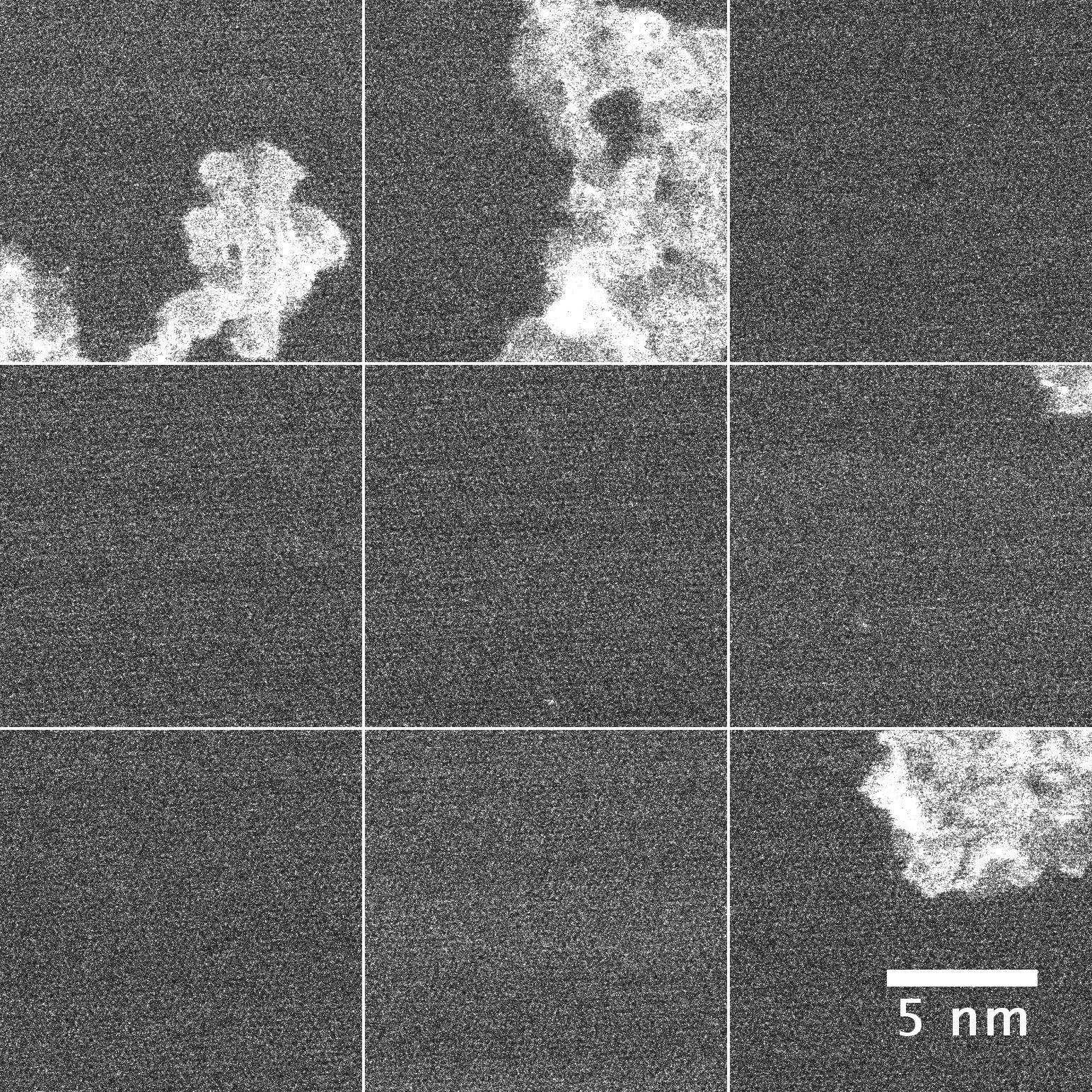}
\includegraphics*[width=0.5\linewidth]{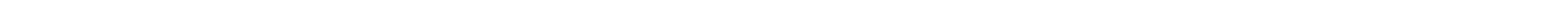}
\includegraphics*[width=0.5\linewidth]{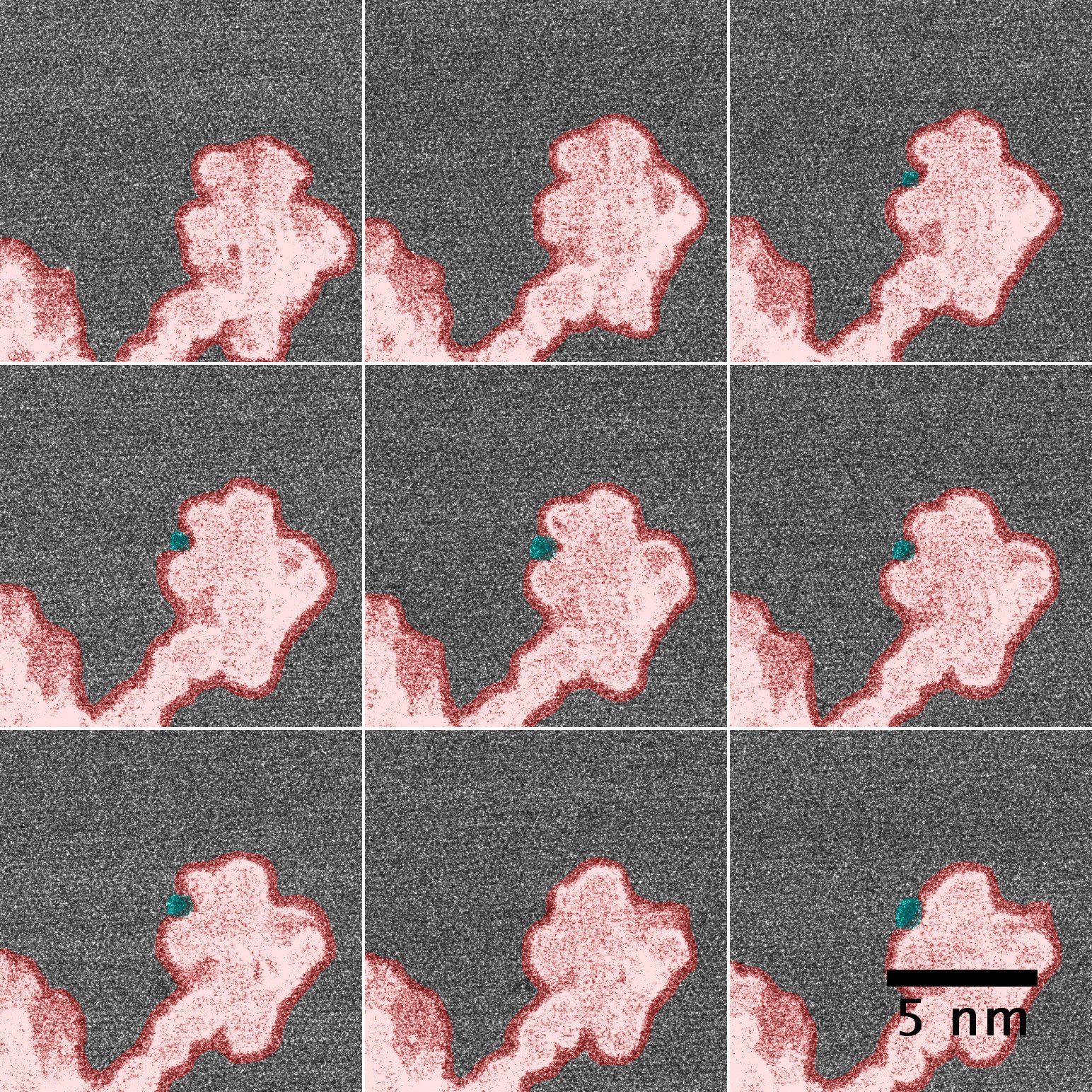}
\caption{%
  \textit{top grid}: Low dose snap-shots of consecutive, spatially separated sites.
  \textit{bottom grid}: The first and then every tenth low-dose aquisition at the same site. A dynamic hole forms at the central beam parking position. The lattice is barely visible. Point defects are not directly visible. Contaminated areas \add{(red)} and holes \add{(cyan)} are excluded from data analysis.}
\label{raw_map}
\end{figure}

All raw STEM images were taken at a Nion Ultra STEM 100, operated at 100~kV, and with the medium-angle annular dark field (MAADF) detector spanning a range of ca. 60-200~mrad. \add{We used a commercial graphenea\textsuperscript{\textregistered} specimen of graphene on a quantifoil\textsuperscript{\textregistered} TEM grid.} The 12~nm field of view was scanned with 2048x2048 pixels at a dwell time of only $0.5$~$\mu$s. We employed a user written extension of the microscope control software (Nion Swift) to facilitate the task of automatically mapping areas on the order of $\mu m^2$ \cite{Mittelberger2017}, without prior exposure of the pristine sample for focusing or stigmation. For the experiment reported here, the microscope was intentionally operated at 100~kV, so that knock-on damage leads to the formation of vacancies \cite{Meyer2012a}. We also intentionally obtained multiple micrographs on each sample location, so as to obtain beam-induced defects in subsequent exposures. The acquisitions were repeated until either the clean graphene coverage dropped below 30\%, which typically happened after 30 frames, or to a maximum of 80 frames. A total of 1187 frames with a 12\,nm field of view were recorded, each with a dose of $1.8\cdot10^4\,e^-/$\AA$^2$. The post-processed signal to noise ratio of $-6.9$~dB is measured in accordance to Ref\@.~\cite{Kramberger2016} as the contrast of the graphene lattice compared to the noise level. Examples are shown in Fig\@.~\ref{raw_map}.  Per frame, this dose is more than 300 times lower than the dose reported in Ref.~\cite{Krivanek2010d} for MAADF imaging of single-layer light element samples, and even the cumulative dose of 30-80 exposures is less than the typical dose for an individual, high signal to noise ratio MAADF image of graphene. \add{In this context we also point out again that each acquisition series starts from zero exposure, after focusing elsewhere, so that we record a signal starting with the first electron on a fresh spot of the sample. This is in contrast to conventional, manual imaging of materials, where typically a region of interest is exposed to high doses before even capturing any data}.

\section{Data processing and analysis}

As first step of the data processing, areas containing bulk contaminations or holes in the graphene are masked out by a thresholded median filter followed by a dilation step \add{(an example of masked out areas is given in the lower grid in Fig. 1)}. The micrographs were also median filtered with a 3x3 kernel, to eliminate outliers. \del{The detailed settings for different aqusition batches are tuned manually with visual feedback.}

The next step is to identify the graphene lattice in every low-dose micrograph with atomic precision. In total, we consider six parameters that allow us to bring the lattices of all exposures into registry: The first two are the  direction and scaling of a linear distortion, the next two parameters describe the rotation and lattice constant of the graphene lattice, and lastly, two cartesian coordinates are used to define a translational reference point.

\begin{figure}[htb]%
\includegraphics*[width=0.5\linewidth]{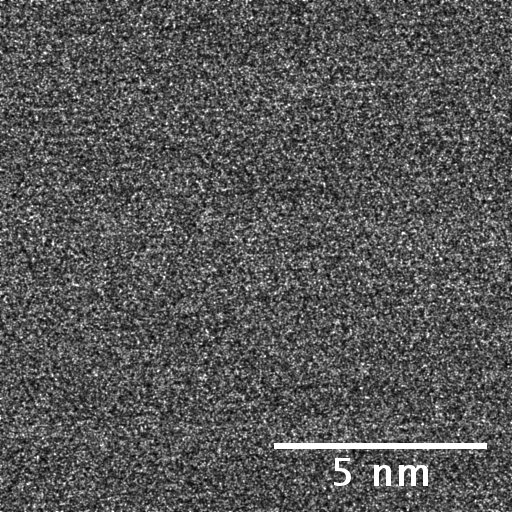}\includegraphics*[width=0.5\linewidth]{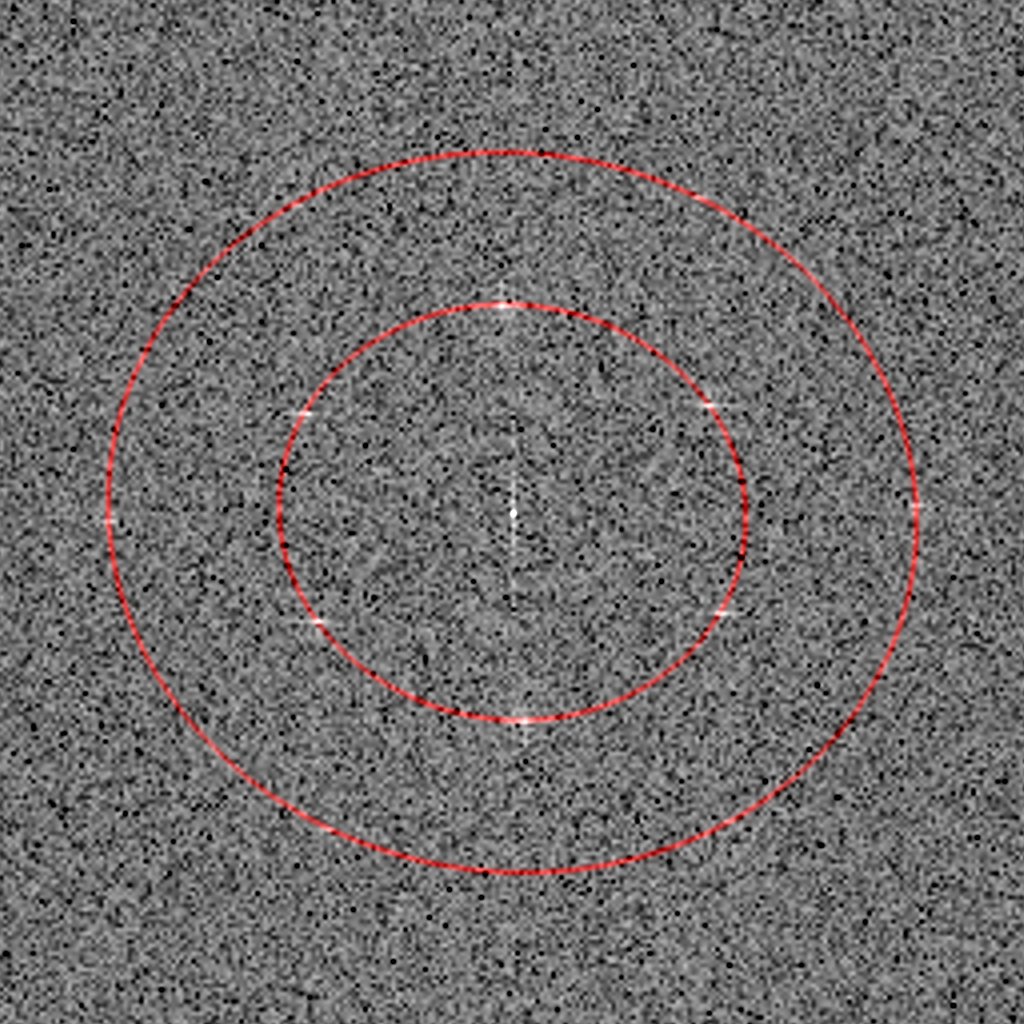}
\caption{%
  \textit{left panel}: A representative low-dose micrograph. \textit{right panel}: Central region of the power spectrum of the Fourier transform with elliptical fits for the first and second order spots.}
\label{raw_fft}
\end{figure}

Figure~\ref{raw_fft} shows a representative low-dose scanning transmission electron microscopy (STEM) micrograph of an (indiscernible) clean freestanding graphene sheet. The most straightforward way to identify the lattice for the alignment of low-dose micrographs is the power spectrum of the discrete Fourier transform (FT). Its central part is also displayed in Fig\@.~\ref{raw_fft}. The bright spots in the power spectrum are the fingerprint of the graphene lattice. The elliptic fits contain \add{convoluted information about the scanning distortions and sample tilts.} The first four parameters for the hexagonal sampling are initialized from elliptical fits like the one in Fig\@.~\ref{raw_fft}. We then optimize all six parameters using a comparison in real space, which does not imply periodic boundary conditions and can also be applied to irregularly shaped regions of clean graphene. For this purpose, a simple coordinate descent along the six parameters is used.

\begin{figure}[htb]%
\includegraphics*[width=\linewidth]{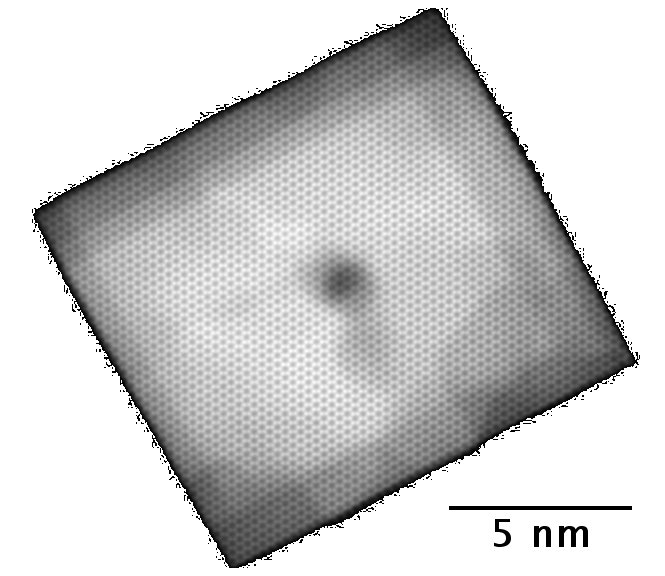}
\caption{%
  Overlay of 1187 aligned and resampled micrographs. The long range contrast variations arise from masking out irregular holes and areas with thick contaminations.}
\label{oview}
\end{figure}

The data is then resampled onto hexagonal pixels, as described \add{in full detail earlier} \cite{Kramberger2016}. The \add{symmetry} \del{choice} of hexagonal pixels as well as hexagonal unitcells allows to represent all point symmetries (rotations and mirrors) of the graphene lattice without any interpolation. We use a sampling density of 12 hexagons per carbon-carbon bondlength for the averaged unitcells during image alignment optimization. The final data for the maximum-likelihood reconstruction is then obtained by another re-sampling with 4 hexagons per \add{C-C} bondlength. \add{During resampling, we also correct the sample tilts, scanning distortions and scaling. For the re-sampling to hexagonal pixels and also for the removal of outliers via median filtering as described above, it is useful that the initial data is highly oversampled.  The hexagonal sampling results in a pixelarea of $0.11$\,\AA$^2$, well suited for atomic resolution.} Figure~\ref{oview} shows the graphene lattice in the average of all aligned images. This averaging does reveal the central parking spot of the electron beam, but it cannot reveal any point defects.

Besides well aligned input data, the maximum likelihood algorithm requires a noise model, i.e., an expression for the probability of measuring a value $k$ when assuming a mean value $\lambda$. Since the experimental detector response is difficult to model, we extract the PDF \add{directly} from the experimental data. For this purpose, the extracted data is grouped into bins by the absolute contrast (as defined by the standard deviation) in the translationally averaged unitcell. A logarithmic grouping with 10-15\% wide bins is found to be adequate for typical low-dose noise levels.

\begin{figure}[htb]%
\includegraphics*[width=\linewidth]{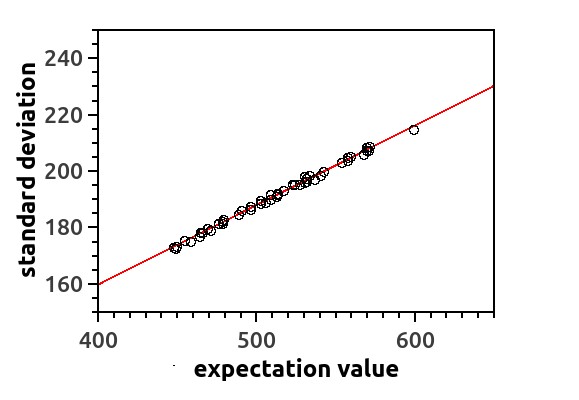}
\includegraphics*[width=\linewidth]{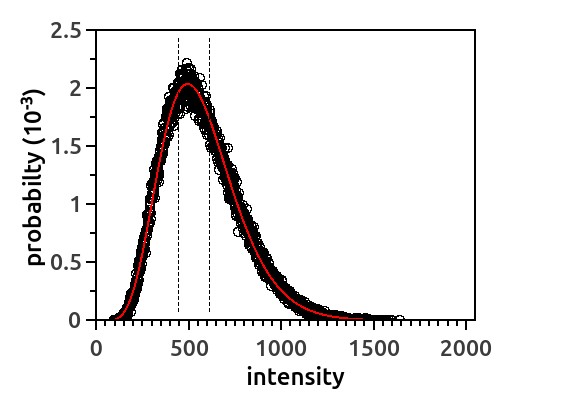}
\caption{%
  \textit{top panel:} linear correlation between the expectation value and the standard deviation in equivalent pixels. \textit{bottom panel:} The actual histogramm (black circles) of observed values on equivalent pixels for a typical expectation value of 512. The brightness in noise free graphene varies between 441 and 608 (dashed verticals). The histogramm is approximated by a Gamma distribution (red line).}
\label{contrast}
\end{figure}

The panels in Fig.~\ref{contrast} do establish two crucial empirical facts. Firstly, there is a linear correlation between the expectation value and the standard deviation of a pixel, and secondly the empirical PDF can be represented by a Gamma distribution. \add{This is in striking contrast to the ideal signal that should consist of Poisson distributed integer scattering events per pixel. We attribute this finding to the strong influence of the signal processing hardware (scintillator, photomultiplier, electronic amplifier and analog to digital conversion, etc) on the resulting noise. Regardless of its origin, the experimentally obtained noise model allows} to represent the PDF with a smooth analytic function that can be extended for expectation values below and above the contrast regime of clean graphene. 

The experimental images were split into smaller hexagonal pieces (supercells)  with $3.35\,$nm$^2$ each. This size is chosen to be just large enough to contain our expected defects. \cite{Kramberger2016} Also, the supercells are allowed to overlap by half of their diameter, to ensure that every point defect is captured completely at least once. A total of $9\cdot10^4$ supercells were extracted from the data. With these supercells and the experimentally obtained PDFs, the likelihood of a basis set of model images to account for the entirity of the measured data can be directly calculated as \cite{meyer2014atomic,Kramberger2016}:

\begin{eqnarray}
 L & = & \prod_{f}\sum_{b}\frac{w_{b}}{J}\sum_{j=1}^{J}P_{b,f,j}\label{eqn:likelihood}\\
 P_{b,f,j} & = & \prod_{p}P_f(k_{f,p},\mu_{b p_j})
\end{eqnarray}

Here the index $f$ denotes the different extracted hexagonal supercells, or frames for short; $b$ runs through the different models in the basis; $j$ denotes the space and pointgroup symmetry equivalent configurations of the frame, and $p$ runs over all pixels within the models or frames. With the measure of $L$ the models and their relative weights can be varied to find the most likely explanation for the observed data. The probability $P_f$ to observe a value $k$ for a given expectation value $\mu$ in a model depends on the contrast group of the frame $f$. Trials for changed weights or pixel values can be rejected or accepted to increase the likelihood $L$.

\begin{figure}[htb]%
\includegraphics*[width=\linewidth]{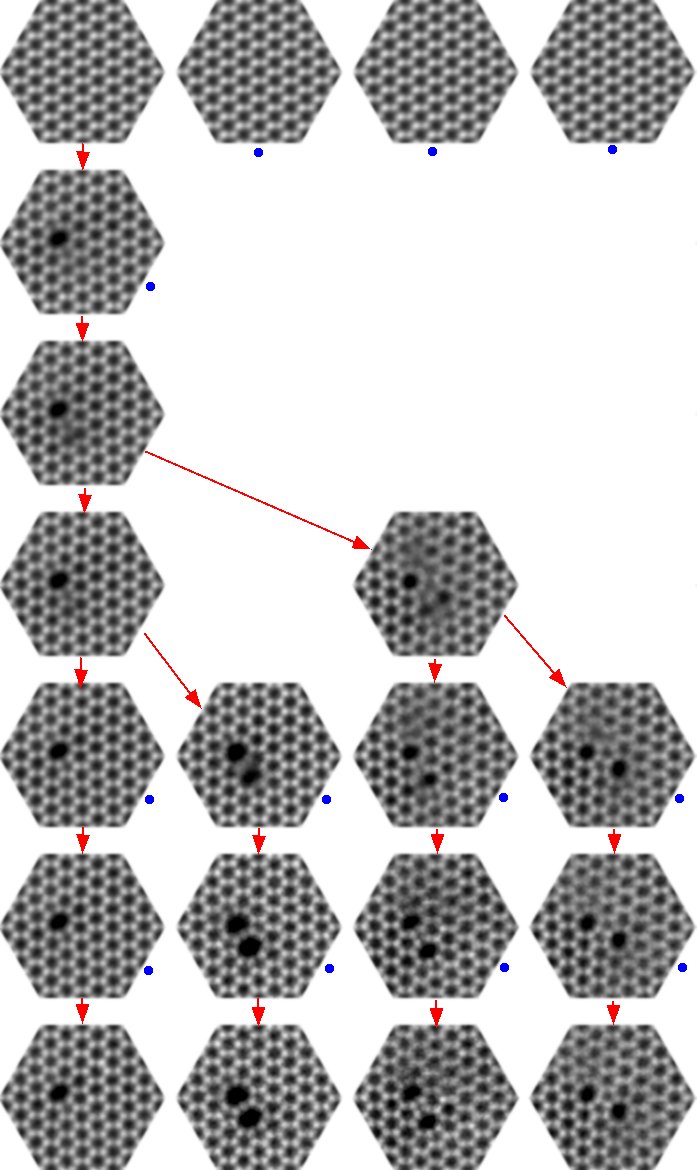}
\caption{ %
The four initial models are in the top row. Descendants are connected with red arrows. Blue dots indicate hidden branches of equivalent models. The bottom row shows a selection from a total of 128 descendants. The weights are from left to right 10.4\%, 0.5\%, 1.8\% and 1.7\%.
}
\label{reconstruction}
\end{figure}

The likelihood maximization is initialized with a set of four identical models representing the empty lattice (top row in Fig.~\ref{reconstruction}). Upon maximization one of them quickly converges to the model in the second row in Fig.~\ref{reconstruction}. After duplicating all models and further optimization of the likelihood one of the descendands of the model in the second row morphes into the one in the third row. The alternating steps of optimization and cloning can be repeated until no more new defects are found. The selection presented here focuses on the direct predecessors of the 4 fundamentally different vacancy defects and disregards the visually similar clones and empty lattices, naturally occuring from this scheme. The red arrows trace the direct lines of descendance. Blue dots mark the roots of entire branches, not included in the selection. The collected weights over equivalent appearances of the four archetypical defects are from left to right 10.4\%, 0.5\%, 1.8\% and 1.7\%. \add{These frequencies correspond approximately to the occurances per model or 3.35\,nm$^2$.} 

\section{Discussion of defect structures}

It is interesting to note that the only divacancy we find in the maximum likelihood reconstruction is the 5-8-5 defect with a weight of $10.4$\% while the other divacancies (those labeled 555777 and 555567777 in Ref\@. \cite{Kotakoski2011d}) are absent. From the lowest confirmed detection we can conclude that the 5-8-5 defect dominates over the other divacancies by at least a factor of 20. This is very different from the more balanced divacancy populations reported from \add{lower voltage and much higher dose aquisition}  series of one single divacancy \cite{Boerner2016,kotakoski2014imaging} \add{(note also that in the earlier work in Ref. \cite{Kramberger2016}, where we used conventional, high-dose experimental data that was artificially resampled to mimic low-dose conditions, the 555777 and 555567777 defects could be recovered)}.

The set of models follows the pattern from the most simple divacancy to paired divacancies and also a five fold vacancy with a bridging atom. This series can be rationalized considering that the primary dynamics is effectively described by the e-beam removing entire bonds \cite{Kotakoski2011b}: After the first atom was kicked out, it becomes very likely to immediately remove one of its neighbors. The absence of single vacancies as well as any other more complex divacancy configurations in Fig\@.~\ref{reconstruction} suggests that under irradiation with 100\,keV electrons, the removal of an entire \add{carbon dimer dominates, while the series of Stone-Wales transformations that would be necessary to form vacancies with three or four pairs of a 5 and a 7 membered ring \cite{Kotakoski2011,Kotakoski2011d} occur only under much higher doses of irradiation.} The model with an bridging atom \cite{Robertson2014} can be obtained by kicking out one of the atoms of the central square in the $3^{rd}$ model. The exclusive observation of this specific high symmetry bridge defect suggests that the ejection of the $5^{th}$ atom does lead to a relatively long lived structure. \add{Single vacancies were not found in this study, but we point out that they are also only very rarely seen in higher-dose TEM or STEM studies of graphene.} 

\add{The lowest demonstrated detection is the one of the 5-fold bridge vacancy with a weight of only $0.5$\%, corresponding to a total of $\sim450$ occurences or a density of one defect per $\sim700\,\textrm{nm}^2$. The 585 di-vacancy with a weight of $10.4$\%, in contrast, should have $\sim9400$ occurences in the data or a density of one defect per $\sim32\,\textrm{nm}^2$.
  At a S/N of -6.9\,dB roughly 100 cases or $\sim0.1$\% would be needed to obtain an satisfactory S/N of $\sim+3.1$\,dB. The algorithm was shown to succeed on extremely low signal levels of $-14$~dB with simulated data\cite{Kramberger2016}. However, for the practical implementation the data processing does require the lattice spots to be detectable in the Fourier transform, which will impose another S/N limit (dependent on the field of view per exposure).  We point out that there is no sharp detection limit in terms of the required defect density, but rather, whether a structure can be recovered depends also the amount of available data, its S/N, and the size and contrast of a feature.}

\section{Conclusions}

We have demonstrated that the retrieval of effective high dose images is feasible for sizeable sets of low-dose annular dark field micrographs. In the current case the individual exposures have an electron dose of $1.8\cdot10^4\,e^-/$\AA$^2$ and a signal to noise ratio after \add{resampling to the targeted resolution} of $-6.9\,$dB. Under these conditions the reconstruction algorithm is successful in independently identifying one missing pair of carbon atoms and the surrounding lattice relaxation as a reoccuring feature in the graphene lattice. The algorithm can also discriminate \del{single and}paired missing bonds as well as a five fold bridge-vacancy. Notably, the populations of extended di-vacancy reconstructions with 3 or 4 pairs of a 5 and a 7 membered ring are found to be below the confirmed detection threshold of 500 occurences and are at least 20 times less frequent. The set of the most frequently occuring vacancy defects can be rationalized in terms of faster atom and atom pair removal by a 100\,keV electron beam as compared to the sequential Stone- Wales transformations, that would be required to form the more extended divacancies. 
\add{The low-dose imaging approach combined with the maximum likelihood reconstruction holds a promise for studying native defects in a material while minimizing changes in the structure caused by the electron irradiation.  The successful image recovery of as little as a few (missing) carbon atoms is a promising sign for more challenging specimen, such as molecules deposited on graphene, or functional groups attached to it.}

\section*{acknowledgement}
We acknowledge support from the European Research Council (ERC) Project
No. 336453-PICOMAT.

\providecommand{\nosc}{}
\let\textsc\nosc
\providecommand{\othercit}{}
\providecommand{\jr}[1]{#1}
\providecommand{\etal}{~et~al.}

\end{document}